\newcommand{\CLASSINPUTtoptextmargin}{0.75in}
\newcommand{\CLASSINPUTbottomtextmargin}{1.01in}
\documentclass[conference]{IEEEtran}
\usepackage{amsmath}
\usepackage{multirow}
\usepackage{multicol}
\usepackage{algpseudocode}
\usepackage[inline]{enumitem}
\usepackage{graphicx}
\usepackage{booktabs}
\usepackage[numbers,sort]{natbib}

\usepackage{soul}
\usepackage{latexsym}
\usepackage{amsmath}
\usepackage{gensymb}
\usepackage{balance}
\usepackage[skip=6pt]{caption}
\usepackage{pifont}
\usepackage{anyfontsize}
\usepackage{subfigure}
\usepackage{url}

\usepackage{etoolbox}

\newcommand{\zerodisplayskips}{%
  \setlength{\abovedisplayskip}{3pt}%
  \setlength{\belowdisplayskip}{3pt}%
  \setlength{\abovedisplayshortskip}{3pt}%
  \setlength{\belowdisplayshortskip}{3pt}}
\appto{\normalsize}{\zerodisplayskips}
\appto{\small}{\zerodisplayskips}
\appto{\footnotesize}{\zerodisplayskips}

\AtBeginDocument{%
  \providecommand\BibTeX{{%
    \normalfont B\kern-0.5em{\scshape i\kern-0.25em b}\kern-0.8em\TeX}}}

\usepackage{subfigure}

\usepackage{enumitem}
\setitemize{topsep=2pt,parsep=0pt,partopsep=0pt,leftmargin=15pt}

\usepackage{algorithm}
\usepackage{algpseudocode}

\title{Split CNN Inference on Networked Microcontrollers \\
\thanks{This work is partially supported by the Strategic Priority Research Program of the Chinese Academy of Sciences (Grant no. 2023YFC3321401) and the EU’s Horizon Europe HarmonicAI project under the HORIZON-MSCA-2022-SE-01 scheme with grant agreement number 101131117. Part of the work was done when B. Li was a visiting PhD student at TU Delft.}
}

\author{}
\author{\IEEEauthorblockN{
		Junyu Lu, Shashwath Suresh, Hao Liu, Qi Hong, Qing Wang
		\IEEEauthorblockA{
		\textit{Delft University of Technology, Delft, The Netherlands}
		}
	}
}

\begin{document}
\maketitle

\begin{abstract}

Running deep neural networks on microcontroller units (MCUs) is severely constrained by limited memory resources. While TinyML techniques reduce model size and computation, they often fail in practice due to excessive peak Random Access Memory (RAM) usage during inference, dominated by intermediate activations. As a result, many models remain infeasible on standalone MCUs. In this work, we present a fine-grained split inference system for networked MCUs that enables collaborative inference of Convolutional Neural Networks (CNN) models across multiple devices. Our key insight is that breaking the memory bottleneck requires splitting inference at sub-layer granularity rather than at layer boundaries. We reinterpret pre-trained models to enable kernel-wise and neuron-wise partitioning, and distribute both model parameters and intermediate activations across multiple MCUs. A lightweight, resource-aware coordinator orchestrates the inference across MCU devices with heterogeneous resources. We implement the proposed system on a real testbed and evaluate it on up to 8 MCUs using MobileNetV2, a representative CNN model. Our experimental results show that CNN models infeasible on a single MCU can be executed across networked MCUs, reducing the per-MCU peak RAM usage while maintaining the practical end-to-end inference latency. All the source code of this work can be found~here:
\url{https://github.com/shashsuresh/split-inference-on-MCUs}

\end{abstract}

\thispagestyle{plain}
\pagestyle{plain}

\section{Introduction}
\label{chp:chapter_1}

Recent advances in Deep Learning (DL) have enabled the increasing deployment of AI at the network edge, supporting applications like sensing and real-time decision making close to data sources. To reduce latency and reliance on the cloud, there is a growing interest in executing DL model inference on resource-constrained embedded devices. Among these, microcontrollers (MCUs) are particularly attractive due to their low cost, low power consumption, and wide usage\footnote{{The annual shipments of MCUs have reached billions; they now account for about 4 out of every 5 processors shipped in the electronics industry \cite{ESourcing2025EdgeAI}.}}.
However, deploying DL models on MCUs is extremely challenging. Compared to edge accelerators or CPUs, MCUs have only tens to hundreds of KB of Random Access Memory (RAM), limited flash memory, low clock frequencies, and often lack floating-point units. These constraints restrict the complexity and size of DL models that can run on a single MCU.

These facts have given rise to the Tiny Machine Learning (TinyML), which adapts ML models to run on MCUs through model optimization. Existing TinyML techniques mainly focus on reducing model parameters and computation. Methods such as quantization~\cite{quant_google,xnor,post_4_bit,affinequant}, pruning~\cite{pruning_deep_compression,pruning_channel_prune,pruning_lottery_ticket}, and neural architecture search~\cite{nas_mnasnet,lin2021mcunetv2,NAS_monte_carlo} have achieved  success.

However, practical deployments reveal a different bottleneck. Even when model weights fit in flash memory, \emph{inference often fails due to peak RAM usage}. Intermediate activations and temporary buffers dominate memory consumption and frequently exceed the available RAM of an MCU. For convolutional and fully connected layers, even one layer can exceed the RAM capacity of an MCU. As a result, many~DL models remain infeasible on standalone MCUs despite aggressive model compression.

At the same time, modern MCU deployments exhibit~two important properties. First, MCUs are \textit{inexpensive and widely available}, making it practical to deploy multiple devices~instead of relying on a single powerful node. Second, MCUs are \textit{increasingly interconnected} through wired or wireless networks as part of larger embedded and IoT systems. Together, these trends motivate us to propose an alternative execution paradigm in this work: \textit{\textbf{performing inference collaboratively across a network of MCUs rather than on a single device.}} 
Instead of executing a full layer on one device, both model parameters and intermediate activations are distributed across several networked MCUs. Each MCU stores and processes only a fraction of the model, greatly reducing peak memory usage per MCU.
However, enabling such collaborative split inference across a network of MCUs has several challenges.

\emph{Challenge 1: Coarse-grained split inference does not work for MCUs.} Most existing split and distributed inference techniques rely on coarse-grained partitioning at layer boundaries and assume runtime support unavailable on MCUs. In practice, such approaches fail because even a single layer may not fit on any individual MCU.
{\textit{To address this challenge,}} we propose a fine-grained split inference mechanism. Computation is partitioned at sub-layer granularity. Convolutional layers are split kernel-wise, and linear layers are split column-wise. This fine-grained design enables inference even when no complete layer fits on a single MCU. 

\emph{Challenge 2: Efficient coordination across heterogeneous MCUs.} In reality, MCUs often differ in memory size, computation speed, and communication bandwidth. Naive evenly splitting of the model will lead to load imbalance and excessive communication overhead, degrading end-to-end model inference latency.
{\textit{To address this challenge}}, in this~work~we design a resource-aware coordination mechanism. Each MCU is assigned a rating that captures its computation, memory, and communication capabilities. We use this rating to allocate workload among the MCUs, which prevents individual devices from becoming bottlenecks. We further integrate system-level optimizations, including layer fusion, quantization, and workflow-level optimizations, to reduce the memory footprint and communication overhead.

By addressing these challenges jointly, we build a practical system for fine-grained split inference on networked MCUs. We implement the system on a testbed consisting of up to 8 MCUs and evaluate it using the representative convolutional neural network MobileNetV2~\cite{Mobilenetv2}. The results show that fine-grained split inference greatly reduces peak memory usage per MCU, enabling inference for models that are \textit{infeasible on a single MCU}, while maintaining practical end-to-end latency and scalability.
We summarize our contributions as follows:

\begin{itemize}
    \item We propose a fine-grained split inference approach that partitions Convolutional Neural Network (CNN) models at sub-layer granularity, enabling both model parameters and intermediate activations to be distributed across multiple MCUs. Using this approach, we enable inference of CNN models--that cannot run on a single MCU--by splitting the inference execution across networked MCUs.
    
    \item We design a split inference system tailored to MCU environments, including model reinterpretation, fine-grained splitting, and a rating-based workload allocation mechanism that accounts for heterogeneous memory, computation, and communication capabilities across MCUs.

    \item We implement the proposed system on a real MCU testbed and evaluate it using MobileNetV2, a representative CNN model. Our experiments show that splitting the inference to multiple MCUs (up to 8 MCUs in the testbed evaluation and up to 120 MCUs in simulation) significantly reduces peak memory usage per device while achieving practical end-to-end inference latency, and that inference time decreases as more MCUs participate, until communication overhead becomes dominant.
    
\end{itemize}

\section{Background and Motivation}
\label{sec_background}

\subsection{TinyML on MCUs}

TinyML targets deploying deep learning models on highly resource-constrained MCUs~\cite{tinyml}. These devices typically offer only tens to hundreds of KB of RAM, limited flash storage, and low clock frequencies. To cope with these constraints, prior work has proposed model-centric optimizations such as quantization, pruning, and knowledge distillation. 
While effective at reducing model parameters and computation, these techniques primarily aim to fit weights into flash memory. In practice, however, inference on MCUs is often limited by \emph{peak RAM usage}. Intermediate activations and temporary buffers dominate memory consumption, rendering many deep models infeasible even after aggressive compression.

\subsection{Split Inference}

Split inference offers an alternative by partitioning computation across devices or between an edge device and a backend server~\cite{neurosurgeon,coacto}. By offloading parts of the computation or intermediate feature maps, split inference effectively extends available memory and computing resources. 
Most existing split inference systems assume at least one powerful node and focus on latency or energy efficiency. Enabling split inference entirely among MCUs poses different challenges, as all devices are severely constrained and communication overhead becomes critical.

\begin{figure}[t]
    \centering
    \vspace{1mm}
    \includegraphics[height=32mm]{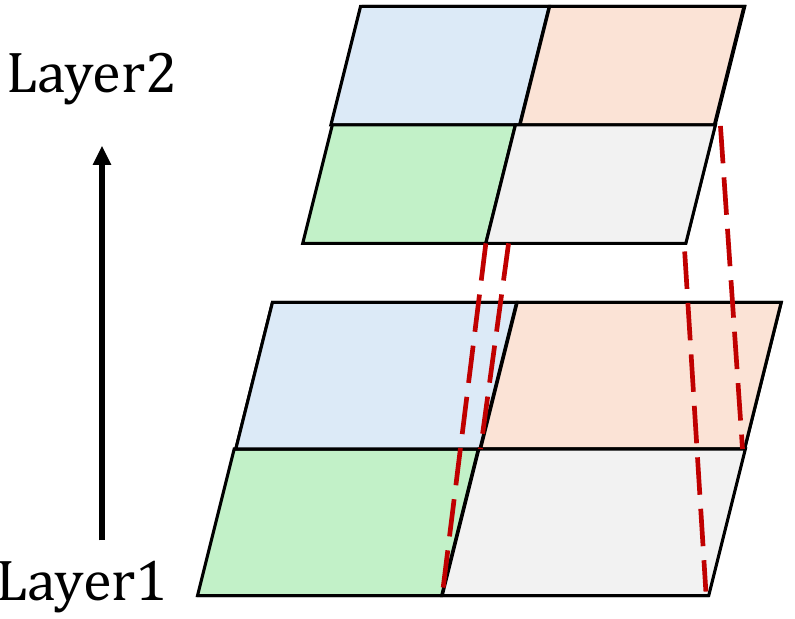} \hfill
    \includegraphics[height=32mm]{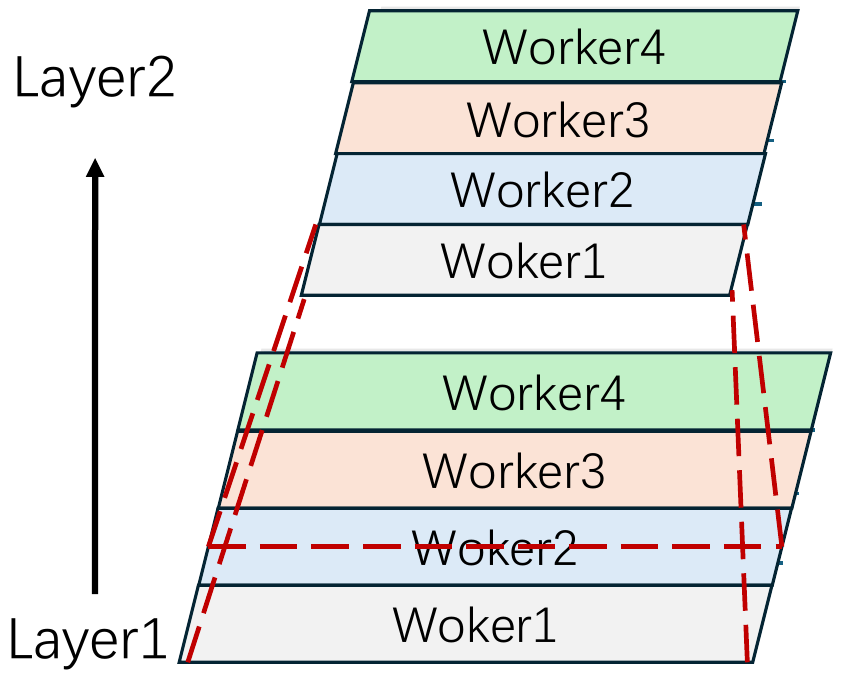} \label{fig:Modnn_split_new}
    \caption{Model partitioning strategy in COTS~\cite{cots} (left) and MoDNN~\cite{MoDnn} (right).}
    \label{fig:cots_split_new}
    \vspace{-4mm}
\end{figure}

\subsection{Model Partitioning Strategies}

The partitioning strategy strongly affects both peak RAM usage and communication cost. Several approaches have been proposed for GPUs and smartphones, but they do not directly translate to MCU environments.

\emph{Spatial Feature Map Partitioning.}
COTS~\cite{cots} partitions intermediate feature maps into spatial sub-matrices and distributes them across workers (Figure~\ref{fig:cots_split_new} (left)). This approach parallelizes computation effectively on GPUs but requires frequent exchange of overlapping activations between workers. Besides, it assumes each worker can store its assigned sub-matrix in memory—an assumption that often fails on MCUs, where intermediate feature maps can exceed available RAM.

\emph{Capability-Aware Spatial Partitioning.}
MoDNN~\cite{MoDnn} assigns uneven spatial partitions based on node capabilities, reducing communication compared to uniform splitting (Figure~\ref{fig:cots_split_new} (right)). However, it still ignores peak RAM usage. Each worker must store all channels of the input and output feature maps, and the full set of convolution kernels. When channel counts are large, this quickly exceeds MCU memory limits. For example, in MobileNetV2, layers with up to 960 channels require approximately 34\,KB of memory for kernel weights alone ($960 \times 3 \times 3 \times 4$ bytes), excluding activations. This exceeds the RAM budget of many power-efficient MCUs.

\subsection{Motivation of This Work}

Existing split strategies are designed primarily to parallelize computation or reduce communication, assuming relatively ample per-device memory. In contrast, MCU-based systems are fundamentally constrained by peak RAM usage. These motivate the need for a split inference strategy that explicitly bounds per-MCU memory at a finer granularity while remaining feasible for highly constrained MCUs.
\section{System Overview}
\label{sec_sys_overview}

\begin{figure*}[t]
    \centering
    \vspace{1mm}
    \includegraphics[width=1.3\columnwidth]{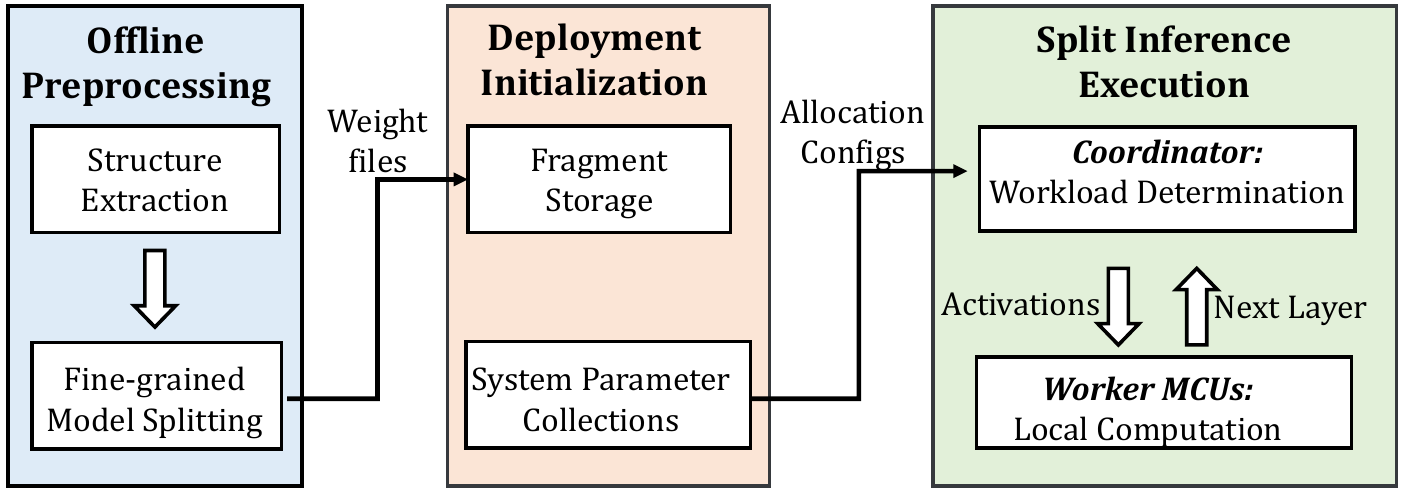}
    \caption{The overall workflow of the proposed fine-grained split CNN inference across networked MCUs.}
    \label{fig_pipleline}
    \vspace{-3mm}
\end{figure*}

We consider a system composed of multiple networked MCUs that collaboratively execute inference for a pre-trained CNN model. The target models are \emph{infeasible} to execute on a single MCU due to memory and/or computational constraints. Rather than relying on aggressive model compression alone, our system enables inference through \emph{fine-grained split execution} across multiple MCUs.

We define two logical roles:
\begin{enumerate*}[label=(\arabic*)]
    \item \textbf{\emph{Coordinator}}, responsible for global orchestration, neuron mapping, activation routing, and synchronization.
    \item \textbf{\emph{Workers}}, hosted on MCUs, which store CNN model fragments and perform local computation.
\end{enumerate*}

A key feature of the system is fine-grained model splitting: instead of assigning entire layers to individual MCUs, computation is partitioned at sub-layer granularity. Depending on deployment constraints, the coordinator may be implemented on a dedicated MCU or an external lightweight device used solely for orchestration. Each worker MCU is assumed to have constrained RAM/memory, limited processing frequency, and heterogeneous networking capabilities.

\textbf{Design Goals.}
We target the following goals: 
{\textit{(1) Bounded peak memory usage}}: Peak memory usage on any MCU~should not exceed physical memory capacity.
\textit{(2) Fine-grained workload distribution}: Both model parameters and intermediate activations are distributable at sub-layer granularity.
\textit{(3) Scalability}: Increasing the number of MCUs should reduce per-device memory and computation load.
\textit{(4) Practical deployability}: The system should operate on bare-metal or lightweight runtimes without requiring heavyweight operating systems.

\textbf{Pipeline.} To meet these design goals, we design the system with a pipeline consisting of three parts: \textit{offline preprocessing}, \textit{deployment initialization}, and \textit{split inference execution}, as shown in Figure~\ref{fig_pipleline}.
\begin{itemize}
    \item \emph{Offline preprocessing.}
The pre-trained model is reinterpreted layer by layer. Structural information such as input and output dimensions, kernel sizes, and receptive fields is extracted. Based on this information, the model is split into multiple \emph{weight fragments} in a \textit{fine-grained way}. Each fragment corresponds to a subset of the model parameters and can be independently deployed to an MCU. 

\item{\textit{Deployment initialization.}}
Each worker MCU receives~its assigned weight fragments and stores them in its flash memory. System parameters such as MCU memory, frequency, and communication capabilities are collected during initialization. These parameters are used
for MCU workload allocation during inference.

\item \emph{Split inference execution.} The designed execution is shown in Figure~\ref{fig_pipleline}. It ensures no MCU needs to store the full set of weights or activations for any layer, thereby bounding the peak RAM memory usage.
The inference execution proceeds in a layer-by-layer manner. For each layer:
\begin{itemize}
    \item The coordinator determines which worker MCUs are responsible for computing specific output neurons.
    \item Required input activations are transmitted to the corresponding workers.
    \item Workers perform local computation using their stored weight fragments.
    \item Partial outputs are returned to the coordinator and forwarded as inputs to the next layer.
\end{itemize}
\end{itemize}

In practical embedded systems, MCUs may differ in computation speed, memory size, and communication capability. To avoid load imbalance, we also introduce a \textit{\textbf{resource-aware coordination mechanism}} for the split inference. Each MCU is assigned a \textit{capability rating} derived from its computation capacity, available memory, and communication characteristics. Workload portions are allocated proportionally to these ratings, preventing slower MCUs from becoming bottlenecks and reducing idle waiting time for faster devices.

\section{Fine-Grained Split Inference Mechanism}

\subsection{Model Reinterpretation}

In our proposed fine-grained split inference mechanism, we need explicit access to neuron-level dependencies, which are not directly exposed by standard deep learning frameworks. Existing frameworks typically operate at layer granularity, making it difficult to reason about individual neurons, their receptive fields, and data dependencies. To enable sub-layer partitioning, we reinterpret pre-trained models into an internal representation that explicitly captures these relationships.

\begin{figure}[t]
    \centering
    \vspace{1mm}
    \includegraphics[width=.5\columnwidth]{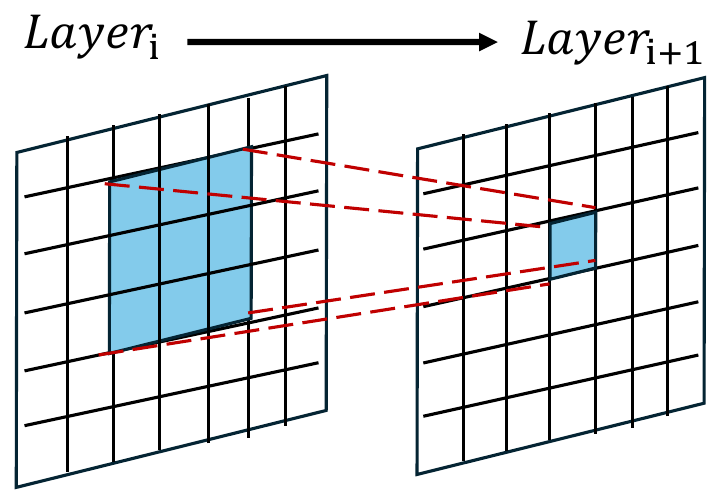}
    \caption{Illustration of the receptive field of an output neuron in a convolutional layer.}
    \label{fig:receptive_field}
\end{figure}

Our reinterpretation process analyzes the model layer by layer and extracts structural metadata, including tensor dimensions, kernel parameters, and weight tensors. For convolutional layers, the receptive field of each output neuron is derived, identifying the exact set of input neurons required for its computation (Figure~\ref{fig:receptive_field}). For linear layers, the mapping between input and output neurons is recorded directly. This information is computed offline and stored as part of the model representation.
In this work, we implement this reinterpretation pipeline using a lightweight inference~framework written in Rust. The model’s computation graph is traced offline, and the extracted metadata and parameters are serialized into a portable representation for deployment.

By making neuron-level dependencies explicit, model reinterpretation provides the foundation for the fine-grained model splitting, activation routing, and memory-bounded execution across networked MCUs.

\subsection{Fine-Grained Splitting Strategy}

Peak memory usage during inference is determined by three components: 
\emph{(i) input activations}, \emph{(ii) weight parameters}, and \emph{(iii) output activations}. 
To bound the peak memory usage on resource-constrained MCUs, we avoid assigning entire layers to individual devices. Instead, we partition computation at the granularity of \emph{output neurons}, enabling both weights and activations to be distributed across MCUs while keeping per-device memory usage bounded.

\begin{algorithm}[t]
\caption{Splitting convolutional layers across workers.}
\label{algo:conv_split_new}
\begin{algorithmic}[1]
\Require $c,h,w$: output channels, height, width of the layer;
$R$: capability ratings; $\mathcal{M}$: set of worker;
$N$: number of workers;
$W$: convolution kernels indexed by channel
\State $s \gets 0$
\For{$r \in \{0,\dots,N-1\}$}
    \State $n \gets \frac{R_r}{\sum_{i=0}^{N-1} R_i} \cdot c \cdot h \cdot w$
    \State $i \gets s$
    \While{$i - s < n$}
        \State $c_1 \gets \left\lfloor \frac{i}{h \cdot w} \right\rfloor$
        \If{$W[c_1]$ not assigned to $\mathcal{M}_r$}
            \State assign $W[c_1]$ to $\mathcal{M}_r$
        \Else
            \State increment usage count of $W[c_1]$ on $\mathcal{M}_r$
        \EndIf
        \State $i \gets i + 1$
    \EndWhile
    \State $s \gets i$
\EndFor
\end{algorithmic}
\end{algorithm}

\begin{figure}[t]
    \centering
    \vspace{2mm}
    \includegraphics[width=\columnwidth]{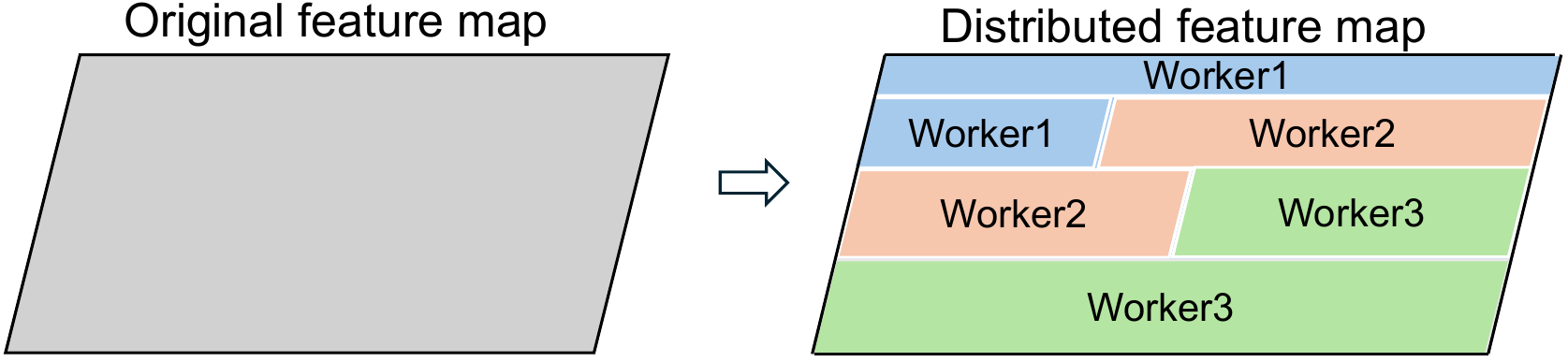}
    \caption{Fine-grained splitting of the convolutional layer across worker MCUs.}
    \label{fig:Split_convolution_new}
\end{figure}

\textbf{Convolutional Layers.}
For convolutional layers, we partition the output feature map into fine-grained patches and assign each patch to a worker MCU, as illustrated in Figure~\ref{fig:Split_convolution_new}.
The splitting procedure is detailed in Algorithm~\ref{algo:conv_split_new}.~The~number of output neurons assigned to a worker, denoted by~$n$,~is first determined based on its capability rating. Then, we~iterate over output positions and identify the corresponding convolution kernel, denoted by $c_1$. If the kernel has not yet been assigned to the worker, the associated weight fragment is distributed; otherwise, the reference count of that kernel is incremented.
This splitting strategy enables highly flexible workload allocation. Different MCUs may be assigned vastly different numbers of output positions, ranging from only a few to several tens of neurons, depending on their capability. Such flexibility allows the system to adapt workload distribution easily and effectively across heterogeneous devices.

\begin{algorithm}[t]
\caption{Splitting linear layers across workers.}
\label{algo:linear_split_new}
\begin{algorithmic}[1]
\Require $h,w$: dimensions of the weight matrix;
$R$: capability ratings; $\mathcal{M}$: set of worker MCUs;
$N$: number of workers;
$W$: weight matrix of size $h \times w$
\State $s \gets 0$
\For{$r \in \{0,\dots,N-1\}$}
    \State $n \gets \frac{R[r]}{\sum_{k=0}^{N-1} R[k]} \cdot w$
    \State $i \gets s$
    \While{$i - s < n$}
        \State assign column $W[:, i]$ to $\mathcal{M}_r$
        \State $i \gets i + 1$
    \EndWhile
    \State $s \gets i$
\EndFor
\end{algorithmic}
\end{algorithm}

\begin{figure}[t]
    \centering
    \vspace{1mm}
    \includegraphics[width=1\columnwidth]{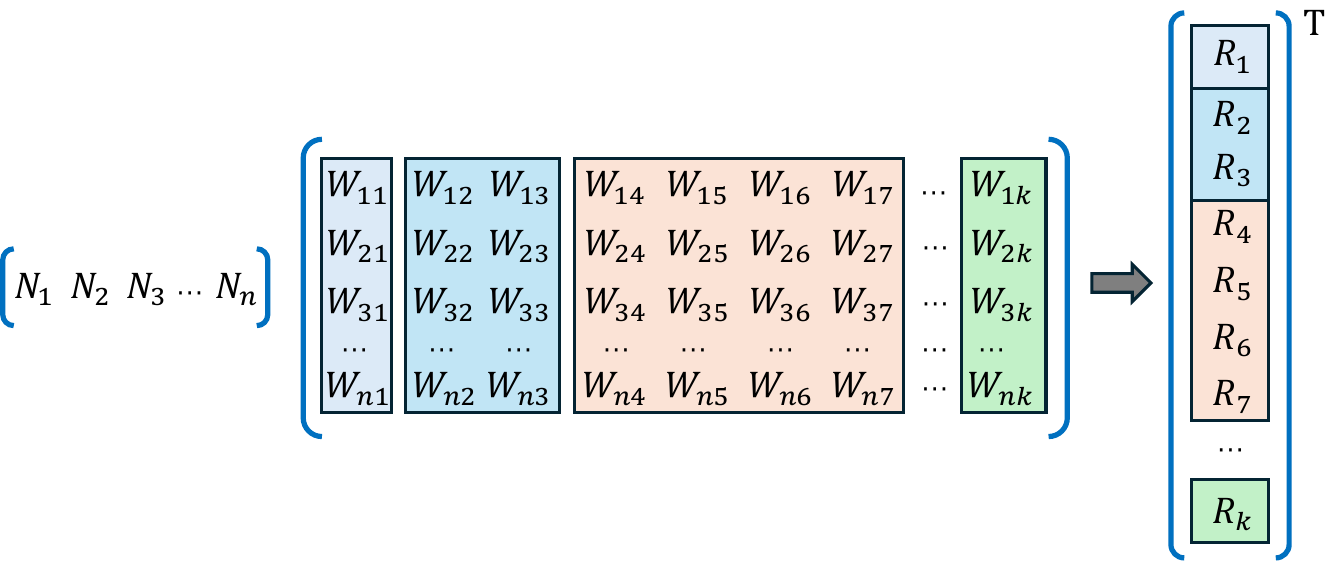}
    \caption{Fine-grained splitting of a linear layer across workers; different colors indicate assignments to different workers.}
    \label{fig:Linear_split_new}
    \vspace{-3mm}
\end{figure}

\textbf{Linear Layers.}
Our splitting strategy of the linear layers is illustrated in Figure~\ref{fig:Linear_split_new} and detailed in Algorithm~\ref{algo:linear_split_new}.
As in convolutional layers, we reuse the same capability ratings $R$ to determine workload allocation across workers. 
However, the weight matrix now has the shape $h,w$, and $W$ represents the weight matrix to be distributed. 
The weight matrix is partitioned at column granularity, where each column corresponds to an output neuron. Similar to the convolutional case, the number of output neurons assigned to each worker is determined by its rating. The key difference is that workload distribution is performed by assigning columns of $W$ to workers, rather than output patches. This column-wise partitioning enables independent computation of output neurons and naturally supports fine-grained splitting.

\subsection{Cross-Layer Activation Mapping}

The coordinator is responsible for routing the intermediate activations produced by layer~$i$ to the workers assigned to compute layer~$i{+}1$. While Algorithms~\ref{algo:conv_split_new} and~\ref{algo:linear_split_new} determine how computation is split \emph{within} each layer, cross-layer execution additionally requires determining \emph{where} intermediate activations must be sent. To this end, we derive explicit cross-layer activation mappings to be used at the coordinator, as detailed in Algorithm~\ref{algo:coordinator_new}.

For each pair of consecutive layers (layers $i$ and $i{+}1$), the coordinator constructs two mappings.  
The \emph{Assignment~Mapping} ($AssignM$) records, for each input activation of layer~$i$, the set of workers in layer~$i{+}1$ that require this activation to compute their assigned output neurons.  
The \emph{Activation Routing Mapping} ($RouteM$) resolves these dependencies by identifying which workers in layer~$i$ must transmit activations to which workers in layer~$i{+}1$ during inference.

The construction proceeds in two stages. In the first stage (Lines~4--20), the coordinator iterates over the output positions of layer~$i{+}1$ and assigns them to workers based on their capability ratings, following the same principle as the convolutional splitting algorithm. For each assigned output position, \texttt{get\_input()} traces its receptive field and identifies the corresponding input activations from layer~$i$. Each such input activation is marked as required by the assigned worker using a bitwise encoding in $AssignM$. This stage captures all neuron-level dependencies between the two layers.

In the second stage (Lines~21--34), the coordinator derives $RouteM$ by iterating over the output activations of layer~$i$ and mapping them to the set of downstream workers that require them, as indicated by $AssignM$. The resulting routing mapping is used at runtime to orchestrate activation transmission between workers.

Algorithm~\ref{algo:coordinator_new} explicitly considers convolutional layers. For linear layers, which typically serve as the classification stage in convolutional neural networks, every output neuron depends on all input neurons. Consequently, all input positions are marked as claimed by the workers assigned to the linear layer, and no receptive-field tracing is required.

\begin{figure}[t]
    \centering
    \vspace{1mm}
    \includegraphics[width=0.6\columnwidth]{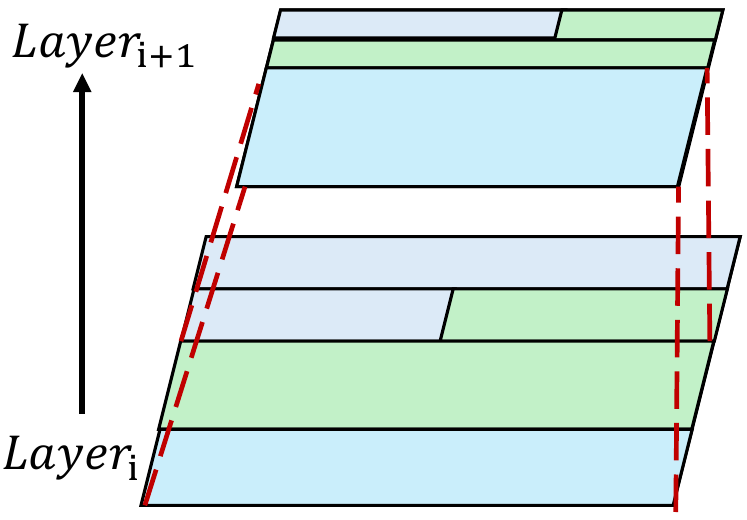}
    \caption{Illustration of the cross-layer mapping.}
    \label{fig:mapping_new}
\end{figure}

Figure~\ref{fig:mapping_new} illustrates the resulting cross-layer activation mapping. In the example, computing the blue region in layer~$i{+}1$ requires a specific subset of activations from layer~$i$, indicated by the red lines. The coordinator maintains this mapping and uses it to route intermediate activations during inference, ensuring correctness while avoiding unnecessary data transfers.

\subsection{Split Inference Execution}

After model partitioning and cross-layer activation mapping are established, inference is executed in a layer-by-layer manner under the coordination of the coordinator. At each layer, workers compute their assigned neurons independently based on the output assignment mapping, while the activation routing mapping determines how intermediate activations are forwarded across layers.

\begin{algorithm}[t] 
\caption{Cross-layer activation mapping for convolutional layers (Coordinator).}
\label{algo:coordinator_new}
\begin{algorithmic}[1]
\Require 
$C_i, H_i, W_i$: channels, height, width of layer $i$; 
$R_i$: capability ratings of worker MCUs for layer $i$;
$N_i$: number of worker MCUs for layer $i$
\Ensure 
$AssignM$: output assignment mapping for layer $i{+}1$;
$RouteM$: activation routing mapping from layer $i$ to $i{+}1$
\State \textbf{Initialize:}
\State \quad $AssignM \gets \mathbf{0}[C_i \times H_i \times W_i]$
\State \quad $RouteM \gets \emptyset$

\vspace{0.3em}
\State \textbf{Stage 1: Construct output assignment mapping}
\State $s \gets 0$
\For{$r \in \{0,\dots,N_{i+1}-1\}$}
    \State $n \gets \frac{R_{i+1}[r]}{\sum_{k=0}^{N_{i+1}-1} R_{i+1}[k]}
              \cdot C_{i+1} \cdot H_{i+1} \cdot W_{i+1}$
    \State $j \gets s$
    \While{$j - s < n$}
        \State $c \gets \left\lfloor \frac{j}{H_{i+1} \cdot W_{i+1}} \right\rfloor$
        \State $h \gets \left\lfloor \frac{\textbf{mod}(j, H_{i+1} \cdot W_{i+1})}{W_{i+1}} \right\rfloor$
        \State $w \gets \textbf{mod}(j, W_{i+1})$
        \State $inputs \gets get\_input(c,h,w)$
        \For{\textbf{each} $p \in inputs$}
            \State $AssignM[p] \gets AssignM[p] \;\mathbf{|}\; (1 \ll r)$
        \EndFor
        \State $j \gets j + 1$
    \EndWhile
    \State $s \gets j$
\EndFor

\vspace{0.3em}
\State \textbf{Stage 2: Derive activation routing mapping}
\State $s \gets 0$
\For{$r \in \{0,\dots,N_i-1\}$}
    \State $n \gets \frac{R_i[r]}{\sum_{k=0}^{N_i-1} R_i[k]}
              \cdot C_i \cdot H_i \cdot W_i$
    \State $j \gets s$
    \While{$j - s < n$}
        \State $c \gets \left\lfloor \frac{j}{H_i \cdot W_i} \right\rfloor$
        \State $h \gets \left\lfloor \frac{\textbf{mod}(j, H_i \cdot W_i)}{W_i} \right\rfloor$
        \State $w \gets \textbf{mod}(j, W_i)$
        \State \textbf{append } $(r, AssignM[c,h,w])$ \textbf{ to } $RouteM$
        \State $j \gets j + 1$
    \EndWhile
    \State $s \gets j$
\EndFor
\end{algorithmic}
\end{algorithm}

\begin{algorithm}[t]
\caption{Split inference execution.}
\label{alg:execution}
\begin{algorithmic}[1]
\Require 
Output assignment mappings $\{AssignM_l\}$;
Activation routing mappings $\{RouteM_l\}$; Input data $x$
\Ensure Inference output $y$

\For{each layer $l$ in the model}
    \State Coordinator determines required input activations for each worker using $RouteM_l$, and sends them to workers
    \ForAll{worker MCU $w$ \textbf{in parallel}}
        \State Receive assigned input activations
        \State $y_w \gets$ Compute assigned neurons with $AssignM_l$
        \State Send partial outputs $y_w$ to coordinator
    \EndFor
    \State Coordinator aggregates partial outputs $\{y_w\}$
    \State $x \gets$ aggregated layer output \Comment{Input for next layer}
\EndFor
\State \textbf{return} $x$
\end{algorithmic}
\end{algorithm}

Algorithm~\ref{alg:execution} details the execution protocol. For each layer, the coordinator identifies the set of input activations required by each worker using the activation routing mapping derived in Algorithm~\ref{algo:coordinator_new}. These activations are transmitted to the corresponding workers, which then perform local computation on their assigned output neurons. Partial results are sent back to the coordinator as soon as the computation completes. The coordinator aggregates these partial outputs and prepares the input activations for the next layer.

Our proposed execution workflow ensures that no worker stores the full activations or weights of any layer, thereby maintaining bounded peak memory usage. At the same time, independent computation across workers allows computation and communication to overlap, improving overall efficiency.

\section{Resource-Aware Workload Allocation}
\label{sec:workflow_optimization}

In networked MCU systems, devices often differ in computation speed, available memory, and communication performance. Even in nominally homogeneous deployments, network latency and contention introduce effective heterogeneity. Uniformly distributing computation across MCUs can therefore lead to load imbalance, where faster devices remain idle while waiting for slower ones, increasing end-to-end inference latency. This section presents a resource-aware workflow optimization strategy that models MCU capabilities, derives capability-based ratings, and incorporates memory constraints to improve the efficiency of split inference.

\subsection{Modeling MCU Capability}
\vspace{-1mm}

For each worker MCU, the total time spent during inference consists of two components:
\begin{enumerate}
    \item \textbf{Computation time}, required to run assigned workload.
    \item \textbf{Communication time}, required to receive input activations and transmit partial results.
\end{enumerate}

Let $W$ denote the assigned workload measured in million cycles. The total execution time $t$ for a worker MCU is:
\begin{equation}
    t = \frac{W}{f} + \left(d + \frac{1}{B}\right) f(W),
    \label{eq:worker_time}
\end{equation}
where $f$ is the MCU clock frequency (in MHz), $d$ is the communication delay per KB, and $B$ is the communication bandwidth (in KB/s).
The function $f(W)$ denotes the amount of data exchanged with the coordinator as a function of workload and is modeled as:
\begin{equation}
    f(W) = K_1 K_c W,
    \label{eq:Workload_to_KB}
\end{equation}
where $K_1$ is a hardware- and algorithm-dependent scaling factor that converts million cycles to KB of output data, and $K_c$ is a \emph{communication coefficient} that captures how much of the computed output must be transmitted. A smaller $K_c$ indicates less communication. For example, when inference runs entirely on a single MCU, $K_c = 0$.

\subsection{Deriving MCU Ratings}

To determine how much workload each MCU can handle, we consider the workload that can be processed within one second. Setting $t = 1$ in Equation~\ref{eq:worker_time} yields:
\begin{equation}
    W\left(\frac{1}{f} + (d + \frac{1}{B})K_1K_c\right) = 1.
\end{equation}

Multiplying both sides by $K_1$ gives:
\begin{equation}
    WK_1 = \frac{fK_1}{(d + \frac{1}{B})fK_1K_c + 1}.
    \label{eq:worker_power}
\end{equation}

The left-hand side represents the volume of output data (in KB) that the MCU can process per second. Based on this observation, we define the \emph{capability rating} $R_i$ of MCU $i$ as:
\begin{equation}
    R_i = \frac{f_iK_1}{(d_i + \frac{1}{B_i})f_iK_1K_{c_i} + 1}.
    \label{eq:rating}
\end{equation}

The rating captures the combined effects of computation speed and communication overhead. In practice, the communication coefficient $K_{c_i}$ depends on the network structure of the CNN and the splitting strategy. It can be estimated through profiling or simulation by measuring the actual volume of data exchanged during inference.

\subsection{Rating-Based Workload Allocation}

Given the ratings $\{R_i\}$, workloads are assigned proportionally to each MCU’s capability. Since model splitting is linear with respect to workload, the size of the weight fragment assigned to MCU $i$ is computed as:
\begin{equation}
    S_i = \frac{R_i S_m}{\sum_{j=1}^{n} R_j},
    \label{eq:weight_size}
    \vspace*{1mm}
\end{equation}
where $S_m$ is the total model size and $n$ is number of MCUs.
This proportional allocation minimizes idle time and aligns workload distribution with heterogeneous device capabilities.

Flash storage and RAM constraints impose hard limits on feasible allocations. After initial assignment, some MCUs may receive weight fragments that exceed their storage capacity. To address this, we adjust ratings while preserving the total rating sum $\sum_{i=1}^{n} R_i$.

For an MCU whose assigned weight size exceeds its storage limit $S_{it}$, the overflowed rating is computed as:
\begin{equation}
    R_{io} = \frac{(S_i - S_{it}) \sum_{j=1}^{n} R_j}{S_m}.
    \label{eq:rating_overflow}
\end{equation}
To avoid excessive load imbalance, the overflowed rating is redistributed evenly among MCUs with remaining storage capacity. This adjustment is repeated iteratively until all weight fragments fit within available storage of each MCU.

\subsection{Other System-Level Optimizations}

To further reduce the model size and model inference speed on resource-constrained MCUs, we apply a set of lightweight model-level optimizations that preserve inference accuracy while simplifying computation and memory usage.
First, we apply \emph{layer fusion} to combine commonly adjacent operations into a single computation. In particular, convolution, batch normalization, and ReLU layers are fused into one composite operation by folding the BatchNorm parameters into the convolution weights and biases, followed by in-place activation. This reduces both the number of operations and volume of intermediate activations, leading to lower memory footprint and inference speed without affecting model accuracy.
Second, we apply \emph{quantization} to convert model parameters and activations from 32-bit floating point to 8-bit integers. This meets practical constraints of many MCUs, which lack hardware support for floating-point arithmetic. 

We also optimize the \emph{execution workflow} to reduce the~synchronization overhead. Workers transmit partial results immediately upon completing local computation, enabling the coordinator to overlap communication with computation across layers. This design minimizes idle waiting time and alleviates the performance impact of centralized coordination.

\section{Implementation}
\label{sec:Implementation}

We implement the proposed split inference system for the MobileNetV2 model~\cite{Mobilenetv2}  on off-the-shelf MCUs.

\subsection{Hardware Platform}

\begin{figure}[t]
    \centering
    \vspace{1mm}
    \includegraphics[width=.6\columnwidth]{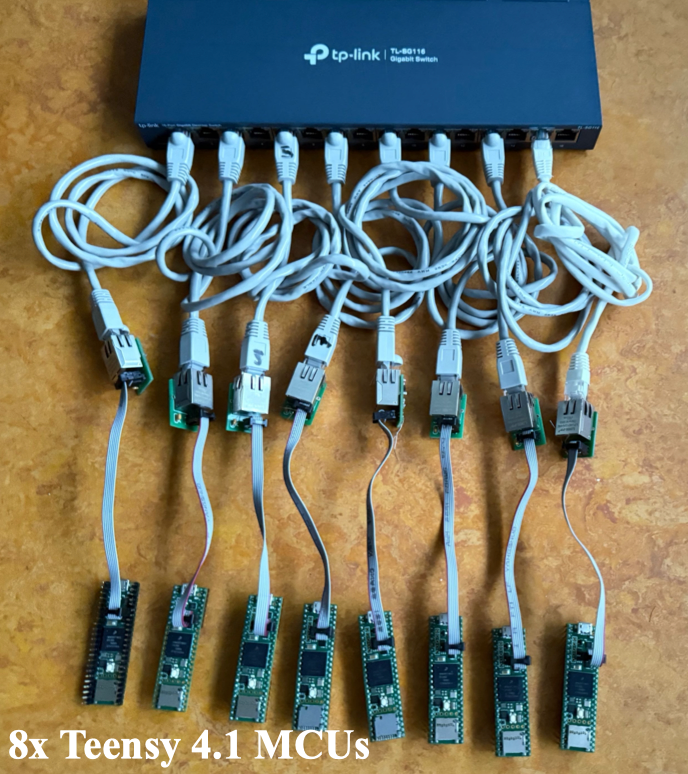}
    \caption{Implementation of the testbed for performance evaluation of our fine-grained split CNN inference on MCUs.}
    \label{fig:testbed}
    \vspace{-3mm}
\end{figure}

We implement the system on the {Teensy~4.1} MCUs~\cite{teensy4.1_web}, each featuring an ARM Cortex-M7 processor running at 600\,MHz, 1\,MB of on-chip RAM, and 8\,MB of flash memory. This configuration satisfies the memory requirements of quantized MobileNetV2 when split across multiple devices while providing sufficient headroom for runtime buffers. Teensy~4.1 also natively supports Ethernet connectivity (up to 100\,Mbps), which simplifies implementation and enables reliable inter-MCU communication.
Our testbed consists of eight Teensy~4.1 MCUs connected via a TP-Link TL-SG116 Gigabit Ethernet switch, as shown in Figure~\ref{fig:testbed}. A PC acts as the coordinator, responsible for orchestration, data logging, and control-plane communication during inference.

\subsection{Software Implementation}

The model preprocessing and workload allocation are performed offline on the coordinator. Based on measured MCU parameters (frequency and communication characteristics), capability ratings are computed and used to split the model into weight fragments. These fragments are serialized into lightweight JSON files and deployed to each MCU’s flash memory using LittleFS, a lightweight file system designed for embedded devices.

During the inference, each worker MCU executes a simple runtime that (i) loads its local weight fragments, (ii) receives input activations from the coordinator, (iii) computes assigned output neurons, and (iv) returns partial results. Tasks are scheduled using a round-robin policy to overlap computation and communication. The complexity of (i) and (iii) largely depends on the algorithm and the hardware used for memory access and output computation, whereas (ii) and (iv) primarily involve data reception and transmission, depending on the communication protocol and channel quality.
The coordinator incurs negligible computational overhead, as it is solely responsible for forwarding intermediate results according to a precomputed mapping stored offline. Overall, the runtime introduces minimal overhead beyond the essential workload.

Communication between the coordinator and workers uses TCP with explicit acknowledgments to ensure reliable transfer under limited buffering. Data is exchanged in fixed-size packets (up to 1400 bytes) to avoid memory pressure on MCUs. 
In the current experimental setup, a single inference pass transfers approximately 4.21 MB of activation data in total across three workers, with the most communication intensive layer requiring around 480 KB per worker.
While it is possible for MCUs to directly forward activations among themselves, we route all intermediate results through the coordinator in the current system to simplify synchronization and measurement. Direct MCU-to-MCU forwarding is left for future work.

\subsection{Automation and Deployment}

We also develop an automation tool to streamline deployment and ensure reproducible experimentation. Given a device role and identifier, the tool automatically configures MCU-specific parameters, distributes the corresponding weight fragments, and flashes the worker application using PlatformIO. Configuration metadata is propagated via environment variables to guarantee consistency across compilation and deployment, followed by a lightweight connectivity check.

This automation enables rapid testbed setup and reconfiguration: adding a new MCU requires only connecting the device and assigning it an ID. For development and debugging, we additionally develop a configuration profile that allows manual override of deployment parameters. Overall, the automation tool significantly reduces deployment overhead and ensures consistent, error-free setup across experiments.

\section{Performance Evaluation}

In this section we evaluate the performance of our proposed fine-grained split inference on networked MCUs. 

\subsection{Experimental Setup}

\emph{\textbf{Platform.}}
We primarily evaluate the system on the testbed, which consists of 8 Teensy~4.1 MCUs. To study the scalability performance, we additionally develop a simulator that preserves the same execution and communication logic while scaling the number of MCUs up to 120.

\emph{\textbf{Model.}}
We use MobileNetV2~\cite{Mobilenetv2} as a representative deep neural network. MobileNetV2 is \emph{infeasible} to execute on a single MCU (e.g., Teensy~4.1) due to the memory constraints. The input resolution is set to $112\times112\times3$, reflecting realistic embedded inference workloads.

\emph{\textbf{Metrics.}}
We consider three metrics: 
(i) \emph{end-to-end inference latency}, measured from input availability to final output; 
(ii) \emph{communication overhead}, measured by transmission latency; and 
(iii) \emph{peak memory usage}, measured as peak heap usage per MCU during inference. 
Peak memory is measured on-device using a lightweight runtime probe, sampled every 1\,ms and at layer boundaries to capture short-lived peaks.

\begin{table}[b]
\centering
\vspace{-3mm}
\caption{$K_1$ under different settings.}
\label{tab:K_1}
\resizebox{.85\linewidth}{!}{
\begin{tabular}{@{}cccc@{}}
\toprule
\textbf{Frequency}& \textbf{Workload} &\textbf{ Time}    & \textbf{$K_1$ }          \\ 
\textbf{(MHz) }& \textbf{(KB)} &\textbf{ (second)}    & \textbf{ (KB/MCycles) }          \\ \midrule
600            & 510.29       & 6.41$\pm$0.3 
& 0.133$\pm$0.007 \\
450            & 510.29        & 7.52$\pm$0.3 & 0.150$\pm$0.006 \\
150            & 510.29        & 16.11$\pm$0.3 & 0.211$\pm$0.004 \\
600            & 421.50       & 5.51$\pm$0.3  & 0.127$\pm$0.007 \\
450            & 421.50       & 6.21$\pm$0.3  & 0.151$\pm$0.007 \\
150            & 421.50       & 13.80$\pm$0.3 & 0.204$\pm$0.005 \\
600            & 730.39      & 7.40$\pm$0.1  & 0.165$\pm$0.002 \\
450            & 730.39       & 9.06$\pm$0.2  & 0.179$\pm$0.004 \\
150            & 730.39       & 21.34$\pm$0.5 & 0.228$\pm$0.006 \\ \bottomrule
\end{tabular}}
\end{table}

\emph{\textbf{Determining $K_1$.}}
We empirically determine $K_1$ for Eq.~\eqref{eq:rating} to derive MCU ratings. We measure $K_1$ by running inference with dummy inputs (no inter-MCU transfers) and timing pure computation at different clock frequencies. On Teensy~4.1, we obtain $K_1$ in the range $[0.127,0.228]$\,KB/MCycle across the tested workloads and frequencies (Table~\ref{tab:K_1}); for the evaluation settings used in this work, the measured $K_1$ values are,~e.g., $K_1{=}$0.133\,KB/MCycle at 600\,MHz, 0.150\,KB/MCycle at 450\,MHz, and 0.211\,KB/MCycle at 150\,MHz.

\subsection{On-Device Evaluation}

\subsubsection{Deployment Feasibility on Networked MCUs}

\begin{figure}[t]
    \centering
    \vspace{1mm}
    \includegraphics[width=.8\columnwidth]{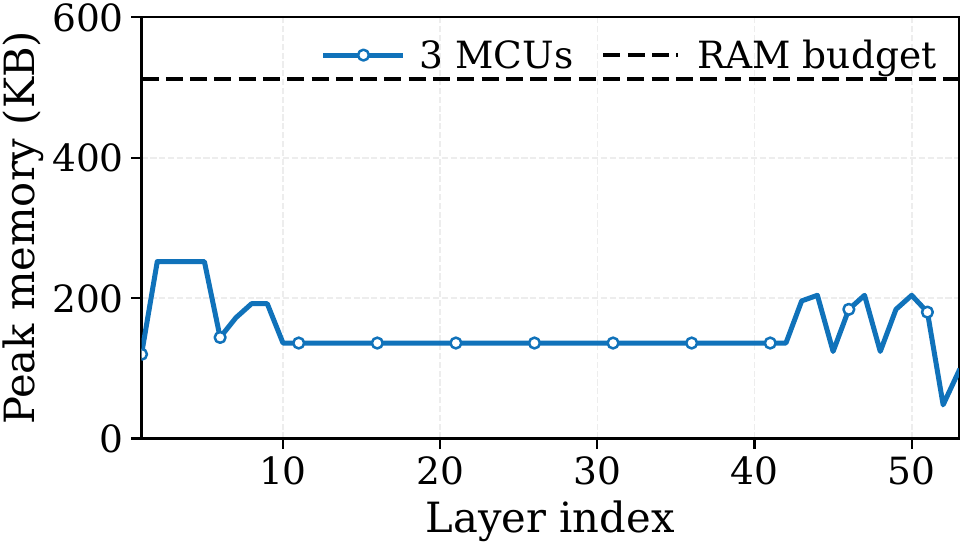}
    \caption{Layer-wise peak RAM usage with 3 MCUs. }
    \label{fig:peak_ram_usage}
\end{figure}

We first validate the deployment feasibility of our proposed fine-grained split inference on networked MCUs. MobileNetV2 has a peak memory footprint of several MB, far exceeding the typical 512\,KB on-chip RAM of a single MCU, leading to out-of-memory failures. Using our designed split inference, we can split the inference of the model across several MCUs.

Figure~\ref{fig:peak_ram_usage} shows the result when the model is split across 3 MCUs. We can observe that the peak RAM usage of every layer remains within the memory budget on each MCU,~confirming that our fine-grained split inference can enable model inference across networked MCUs that is otherwise infeasible. Higher peak memory appears in early layers, consistent with MobileNetV2’s architecture, which produces large intermediate feature maps during initial feature extraction.

\begin{table}[t]
    \centering
    \caption{Performance comparison of distribution strategies (3 MCUs)}
    \label{tab:performance_comparison}
    \resizebox{\linewidth}{!}{
    \begin{tabular}{c c c c c c}
        \toprule
        \multirow{2}{*}{\textbf{Case}} & \multicolumn{2}{c}{\textbf{Configuration}} & \multicolumn{3}{c}{\textbf{Execution Time (s)}} \\
        \cmidrule(lr){2-3} \cmidrule(lr){4-6}
         & \textbf{Freq. (MHz)} & \textbf{Delay (ms)} & \textbf{Evenly} & \textbf{Freq. Only} & \textbf{Optimized} \\
        \midrule
        1 & 600/600/600 & 0/0/0     & 9.80  & 9.80  & \textbf{9.80} \\
        2 & 600/150/450 & 0/0/0     & 20.10 & \textbf{12.40} & 12.52 \\
        3 & 150/396/528 & 0/0/0     & 22.30 & 13.43 & \textbf{13.37} \\
        4 & 450/396/528 & 0/0/0     & 11.44 & 10.75 & \textbf{10.61} \\
        \addlinespace 
        5 & 600/150/450 & 10/0/5    & 32.81 & 33.01 & \textbf{31.50} \\
        6 & 450/396/528 & 20/7/13   & 54.73 & 54.20 & \textbf{47.41} \\
        7 & 600/396/150 & 20/5/10   & 53.08 & 54.83 & \textbf{44.45} \\
        8 & 600/600/600 & 10/20/5   & 49.18 & 49.18 & \textbf{41.95} \\
        \bottomrule
    \end{tabular}
    }
\end{table}

\subsubsection{Effectiveness of Our Rating-Based Allocation}
Next,~we evaluate the effectiveness of our rating-based allocation by comparing it with two baselines: \emph{Evenly} (uniform split) and \emph{Freq.-only} (split proportional to MCU frequency). Each experiment executes one end-to-end inference on 3 MCUs.
To capture heterogeneous conditions, we vary both computation speed and communication delay. For computation heterogeneity, we select MCU frequencies from those supported by the platform, including homogeneous (e.g., 600/600/600\,MHz) and heterogeneous configurations (e.g., 600/150/450\,MHz). For communication heterogeneity, we inject asymmetric delays before send/receive operations, sweeping from 1–20\,ms to emulate low-, medium-, and high-latency links. The details are given in Table~\ref{tab:performance_comparison}.
Cases~1–4 vary frequency only, while Cases~5–8 vary both frequency and delay.

We have three observations from the results shown in~Table~\ref{tab:performance_comparison}.
First, uniform allocation consistently performs worst, as faster MCUs idle while waiting for slower ones. Second, when communication delay is negligible (Cases~1–4), the optimized scheme closely matches the Freq.-only baseline, since computation dominates. Third, once communication delay is introduced (Cases~5–8), the optimized scheme consistently achieves the lowest latency by accounting for both computation and communication. For example, in Case~7, total latency is reduced from 54.83\,s to 44.45\,s, an 18.9\% improvement. These results confirm that incorporating communication costs into workload allocation is essential in realistic heterogeneous deployments.

\subsubsection{End-to-End Inference Latency}

\begin{figure}[t]
    \centering
    \vspace{2mm}
    \includegraphics[width=.75\columnwidth]{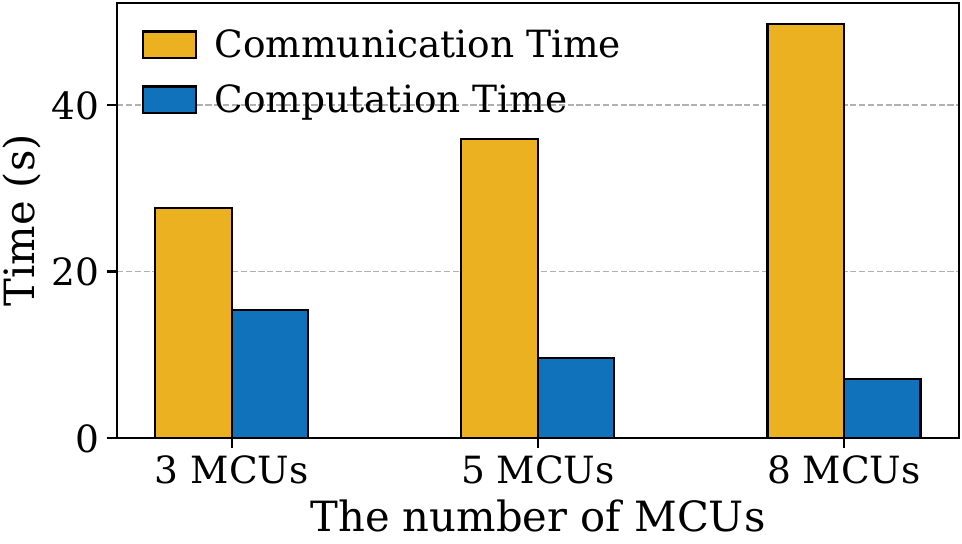}
    \caption{Communication and computation time per inference on 3/5/8 MCUs.}
    \label{fig_time_distribution}
    \vspace{-4mm}
\end{figure}

Figure~\ref{fig_time_distribution} decomposes inference latency into computation and communication as the testbed scales from 3 to 5 to 8 MCUs. The total inference time increases moderately (42.97\,s, 45.61\,s, and 56.89\,s),~driven mainly by communication overhead, which rises from 27.60\,s to 49.82\,s. In contrast, the computation time decreases monotonically (15.37\,s to 7.07\,s), indicating the effectiveness of our workload partitioning.

These results highlight a fundamental trade-off: computation scales favorably with additional MCUs, while communication becomes the dominant cost. Although communication could be further optimized (e.g., batching/pipelining), such optimizations are orthogonal to this work. Our goal is to show the feasibility and effectiveness of fine-grained split inference under a straightforward communication baseline.

\subsection{Performance Breakdown}

\subsubsection{Communication Overhead}

\begin{figure}[t]
    \centering
    \vspace{2mm}
    \includegraphics[width=.75\columnwidth]{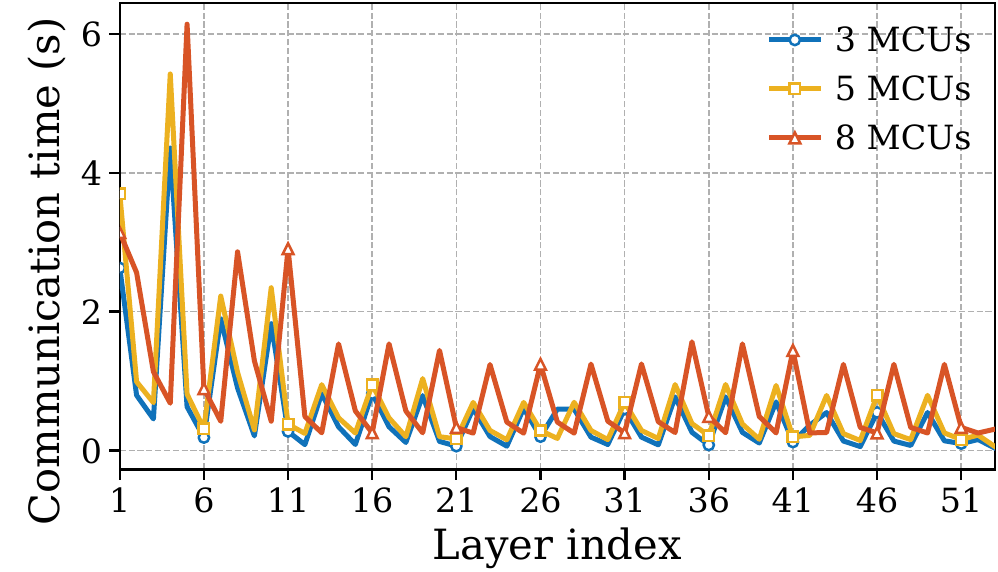}
    \vspace{-1mm}
    \caption{Layer-wise communication time for 3/5/8 MCUs.}
    \label{fig:comm_overhead}
    \vspace{-2mm}
\end{figure}

Figure~\ref{fig:comm_overhead} reports layer-wise communication overhead for deployments on 3, 5, and 8 MCUs. Communication cost increases monotonically with the number of MCUs, as finer partitioning requires more inter-device data transfers. Most communication overhead concentrates in early layers, where intermediate feature maps are largest. This trend explains the growing dominance of communication in larger deployments.

\subsubsection{Computation Scaling}

\begin{figure}[t]
    \centering
    \includegraphics[width=.75\columnwidth]{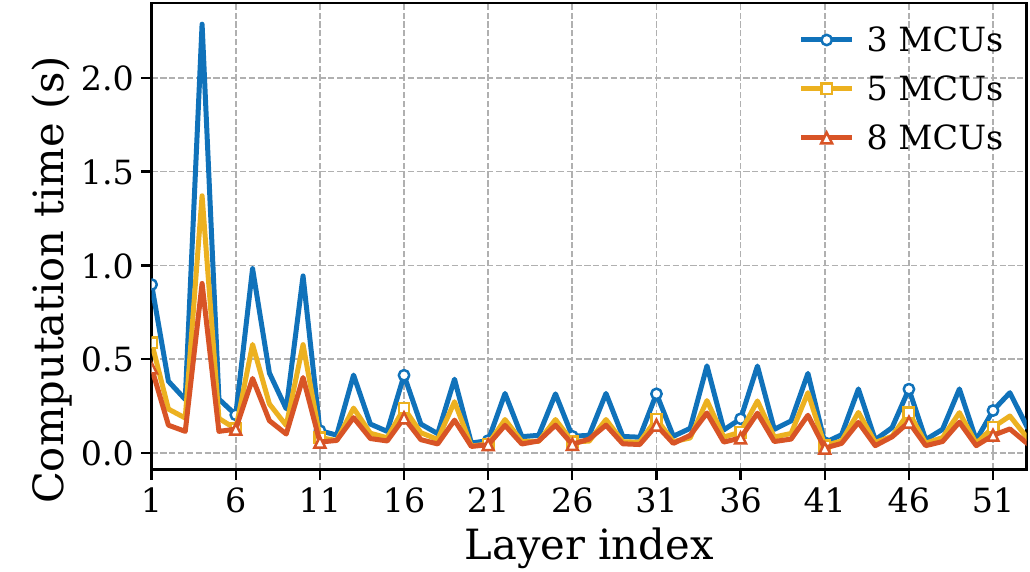}
    \vspace{-1mm}
    \caption{Layer-wise computation time for 3/5/8 MCUs.}
    \label{fig_comp_overhead}
    \vspace{-4mm}
\end{figure}

Figure~\ref{fig_comp_overhead} shows layer-wise computation time across different MCU counts. Computation time decreases consistently as more MCUs are added, with the largest reductions occurring in compute-intensive layers. This confirms that fine-grained partitioning effectively distributes computation and scales the compute component of inference.

\subsection{Scalability Evaluation (Simulation)}

\begin{figure}[t]
    \centering
    \vspace{2mm}
    \includegraphics[width=.75\columnwidth]{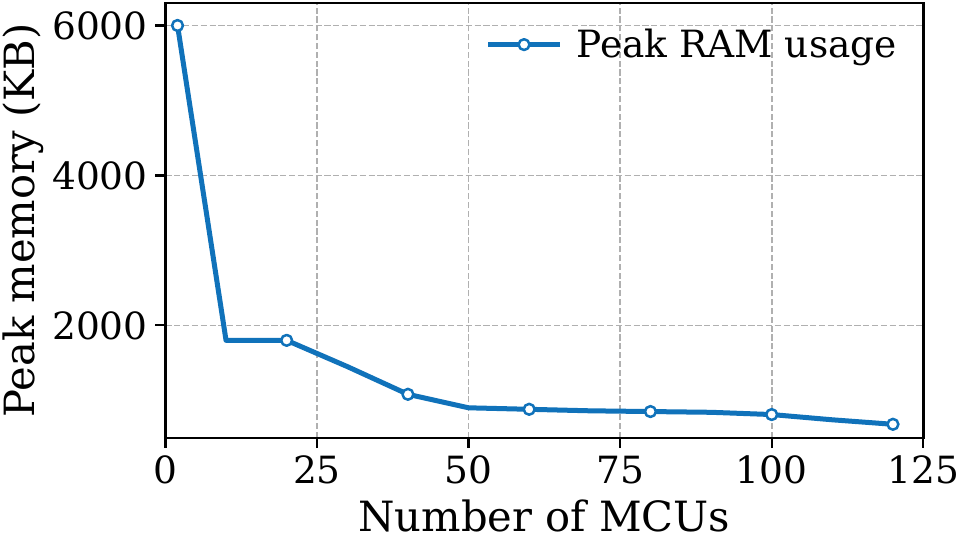}
    \caption{Per-MCU peak memory versus the number of MCUs.}
    \label{fig_120_MCU}
    \vspace{-3mm}
\end{figure}

Finally, we study the memory scalability using simulation, which is impractical to evaluate using real hardware at large scale. The simulator models one coordinator and $N$ worker MCUs using threads, preserving the same partitioning and communication logic.
Figure~\ref{fig_120_MCU} shows peak per-MCU memory usage as the number of MCUs increases. A single MCU exceeds the memory budget and cannot run the model. As the number of MCUs increases, peak per-MCU memory drops by multiple times, making inference feasible on resource-constrained MCUs. The largest gains occur within the first few MCUs, with diminishing returns beyond roughly 10–20 devices. Extending to 120 MCUs reveals a clear saturation trend, quantifying the memory–parallelism trade-off and explaining why small-scale deployments (e.g., less than ten MCUs) are already effective.
\section{Related Work}

{\textbf{TinyML Optimizations for Edge Devices.}} To mitigate the gap between high computational demands and limited hardware resource, extensive research has focused on model compression and efficient architecture design. Model compression techniques, particularly quantization and pruning, serve as the fundamental optimization strategies. Quantization~\cite{quant_multiscale,quant_google,quant_4bit,wang2023new} reduces memory footprint and transmission overhead by representing weights and activations with lower precision. Network pruning~\cite{pruning_global,pruning_splittable,pruning_hardwarefriendly,pruning_adaptive_channel} removes redundant connections or neurons to accelerate inference. Beyond static compression, hardware-aware neural architecture search, such as MCUNet~\cite{lin2021mcunetv2}, automates the design of efficient network topologies tailored for the strict constraints of MCUs. At the system level, lightweight inference frameworks such as TensorFlow Lite for Microcontrollers~\cite{tensorflow_micro} and CMSIS-NN~\cite{cmsis_nn} optimize kernel implementations to minimize latency. However, although they offer a valuable way to optimize the inference, the deployment of large-scale DNNs remains bounded by the physical limitations of a single MCU. This bottleneck necessitates a shift from single-device optimization to a distributed inference strategy. 

{\textbf{Distributed Inference for Edge Devices.}} Model partitioning enables distributed inference by splitting DNN computations across multiple devices. Neurosurgeon~\cite{neurosurgeon} proposes a concept of collaborative inference workflow for edge-cloud servers. CoActo~\cite{coacto} improves the model splitting on a fine-grained level and introduces concurrency of runtime resources to existing distributed inference. COTS~\cite{cots} distributes the inference workflow into several GPUs for acceleration. MoDNNs~\cite{MoDnn} introduces a way to partition the input feature map into individual nodes for the support of smartphones. However, these splitting strategies mainly target the high-performance backend devices (e.g., cloud servers, GPUs), neglecting the scenarios for MCU clusters. It is infeasible to directly apply these techniques in our scenario due to strict memory constraints.
\section{Conclusion}

In this work we present a fine-grained split inference framework for executing deep neural networks across networked MCUs, addressing the severe resource constraints of individual devices. We design a distributed inference workflow that splits models into weight fragments, coordinates execution through a lightweight coordinator, and performs neuron-level mapping across convolutional and linear layers. We further introduce resource-aware workflow optimizations to improve efficiency while preserving accuracy.  Evaluations on 8 MCUs demonstrate that our approach significantly reduces peak per-MCU RAM usage and enables inference that is otherwise infeasible on a single MCU. Overall, we show the practicality of split inference on networked MCUs and provide a foundation for future extensions toward more efficient communication and broader model classes.

\vspace{4mm}
\small 
\noindent \textbf{Acknowledgment:} This work is partly supported by EU's~Horizon Europe {HarmonicAI} project under the HORIZON-MSCA-2022-SE-01 scheme with the grant agreement number 101131117. 
\vspace{2mm}

\balance
\footnotesize 
\bibliographystyle{IEEEtran}
\bibliography{ref}

\end{document}